\documentstyle[12pt]{article}

\def\be{\begin{equation}}
\def\ee{\end{equation}}
\def\a{\alpha}

\def\s{\sigma}

\def\d{\partial} 
\def\l{\lambda} 
\def\det{\mbox{det}}
\def\ln{\mbox{ln}}
\def\exp{\mbox{exp}}
\def\cos{\mbox{cos}}
\def\sin{\mbox{sin}}
\def\ak{a^{+}}
\def\ck{c^{+}}
\def\ra{\rangle}
\def\la{\langle}
\def\Nt{\tilde{N}}
\def\D{\Delta}
\def\n{\tilde{n}}
\def\r{\rho}

\begin{document}

\begin{center}
{\bf On the universal relations for the prefactors in correlation functions 
of 1D quantum liquids.} 
\end{center}
\vspace{0.2in}
\begin{center}
{\large A.A.Ovchinnikov}
\end{center}

\begin{center}
{\it Institute for Nuclear Research, RAS, Moscow}
\end{center}

\vspace{0.2in}

\begin{abstract}

For various one-dimensional quantum liquids in the framework of the Luttinger model 
(bosonization) we establish the relations between the coefficients before the 
power-law asymptotics of the correlators (prefactors) and the formfactors of the 
corresponding local operators. The derivation of these relations in the framework 
of the bosonization procedure allows to substantiate the prediction for the formfactors 
corresponding to the low-lying particle-hole excitations. We also obtain  
the formulas for the summation over the particle-hole states corresponding to the 
power-law asymptotics of the correlators.

\end{abstract}

\vspace{0.2in}

{\bf 1. Introduction}

\vspace{0.2in}

Calculation of correlation functions in the 1D quantum liquids or 
spin systems remains important problem both from theoretical and 
experimental points of view. Although the predictions of the 
critical exponents corresponding to the power-law decay at large 
distances obtained with the help of the mapping to the Luttinger model
(bosonization) \cite{LP},\cite{H} or conformal field theory \cite{C}, 
\cite{MZ} are available for a long time, the calculation of the constants 
before the asymptotics (prefactors) remains an open problem. 
Recently some progress was achieved in relating of the prefactors 
to the certain formfactors of local operators by means of direct 
calculations in the integrable model \cite{S1} and by means of the 
conformal field theory \cite{G}. Moreover, for the XXZ - quantum spin 
model the behaviour of the formfactors for the low-lying particle-hole 
excitations was found \cite{S2}, which agrees with the predictions of 
the recent papers \cite{G}.  

The arguments of ref.\cite{G} on the particle-hole formfactors for the 
low-lying states are based on the combination of the predictions of the 
conformal field theory and the bosonization technique. However there is 
no derivation of the scaling relations for the lowest formfactors in 
the framework of the bosonization technique, although the particle-hole 
formfactors are obtained in the spirit of the bosonization approach. 
Therefore their results on the particle-hole formfactors are not grounded 
enough. 
It would be successive to obtain the results for both type of the formfactors 
within the same method. 
In their derivation of the particle-hole formfactors there is no proof that 
the constant in front of the factor, which depends on the particle and hole 
momenta (see below), is, in fact, the lowest formfactor i.e. there is no 
proof of the whole formula, which can be obtained only in the framework 
of the bosonization approach.  
Thus it is desirable to derive the scaling relations for the lowest 
formfactors entirely using the bosonization technique without use of the 
conformal field theory. This is the main goal of the present paper. 
The second goal is to point out the summation formulas which can be used 
to calculate the sum over the intermediate states for the correlators 
in the framework of the approach of ref.\cite{S2} to the correlators in the 
integrable (XXZ- spin chain) model.  
 
In section 2 we fix the notations and briefly review the theory of 
Luttinger liquid and the bosonization procedure. In section 3 we explain 
how to derive the scaling relations for the lowest formfactors using the 
bosonization approach. Finally in section 4 we derive the particle-hole 
formfactors for the low-lying states and present the summation formulas 
relevant to the calculation of the correlators.

\vspace{0.2in}

{\bf 2. Bosonization.}

\vspace{0.2in}

Consider the effective low -energy Hamiltonian which build up from 
the fermionic operators ($a_k,~c_k,~k=2\pi n/L,~n\in Z$, L- is the length 
of the chain) 
\[
a^{+}(x)=\frac{1}{\sqrt{L}}\sum_{k}e^{-ikx}a_{k}^{+}, ~~~~~
a^{+}_k=\frac{1}{\sqrt{L}}\int_{0}^{L}dx e^{ikx}a^{+}(x), 
\]
corresponding to the excitations around the right and the 
left Fermi- points and consists of the kinetic energy term and the 
interaction term $H=T+V$ with the coupling constant $\l$:    
\be
H=\sum_{k}k(a_k^{+}a_k-c_k^{+}c_k)+2\pi\l/L\sum_{k,k',q}\ak_ka_{k+q}\ck_{k'}c_{k'-q}.      
\label{ham}
\ee
Defining the operators \cite{ML}
\[
\r_1(p)=\sum_k\ak_{k+p}a_k,~~~~\r_2(p)=\sum_k\ck_{k+p}c_k,
\]
where $|k|,|k+p|<\Lambda$, where $\Lambda$ is some cut-off energy, 
which for the states with the filled Dirac sea have the following 
commutational relations 
\[
\left[\r_1(-p);\r_1(p')\right]=\frac{pL}{2\pi}\delta_{p,p'}~~~~
\left[\r_2(p);\r_2(-p')\right]=\frac{pL}{2\pi}\delta_{p,p'},
\]
one can represent the Hamiltonian in the following form: 
\[
H=\frac{2\pi}{L}\sum_{p>0}\left(\r_1(p)\r_1(-p)+\r_2(-p)\r_2(p)\right)+ 
\l\sum_{p>0}\frac{2\pi}{L}\left(\r_1(p)\r_2(-p)+\r_1(-p)\r_2(p)\right).
\]
To evaluate the correlators in the system of finite length and make the 
connection with the conformal field theory predictions, one can proceed as follows. 
First one can define the lattice fields $n_{1,2}(x)$ with the help of the Fourier 
transform as 
\[
\rho_{1,2}(p)=\int_{0}^{L} dx e^{ipx}n_{1,2}(x),~~~~
n_{1,2}(x)=\frac{1}{L}\sum_p e^{-ipx}\rho_{1,2}(p)
\]
This fields have a physical meaning of the local number of the fermions 
above the Fermi level at the right and the left Fermi points. In terms of this 
fields the Hamiltonian has the following form:
\[
H=2\pi\sum_x\left(\frac{1}{2}(n_1^2(x)+n_2^2(x))+\lambda n_1(x)n_2(x)\right)
\]
Considering  the average distribution of the number of extra particles we obtain
the ground state energy in the form
\[
\Delta E=\frac{2\pi}{L}\left(\frac{1}{2}\left((\D N_1)^2+(\D N_2)^2\right)
+\l\D N_1 \D N_2 \right)
\]
where $\D N_{1,2}$ - are the numbers of additional particles at the two Fermi points. 
One can also rewrite the  ground state energy in the sector with the total number of
particles and the momentum $\D N=\D N_1+\D N_2,~~\D Q=\D N_1-\D N_2$
in such a way that the total Hamiltonian takes the following form (this for was first 
proposed in ref.\cite{H}): 
\be
H=u(\l)\sum_p|p|b_p^{+}b_p+\frac{\pi}{2L}u(\l)\left[\xi(\D N)^2+(1/\xi)(\D Q)^2\right],
\label{finite}
\ee
where the parameters $u(\l)=(1-\l^2)^{1/2}$ and $\xi=((1+\l)/(1-\l))^{1/2}$. 
Next one establishes the commutational relations for the fields $n_{1,2}(x)$: 
\[
\left[ n_{1}(x); n_{1}(y)\right]=-\frac{i}{2\pi}\delta'(x-y)~~~~
\left[ n_{2}(x); n_{2}(y)\right]=\frac{i}{2\pi}\delta'(x-y) 
\]
Then introducing the new variables $\n_{1,2}(x)=\sqrt{2\pi}~ n_{1,2}(x)$,
we have the following density of the Hamiltonian 
\be
H=\frac{1}{2}\left(\n_1(x)\n_1(x)+\n_2(x)\n_2(x)\right)
+ \lambda \n_1(x)\n_2(x).
\label{v}
\ee
We also have the following commutational relations
$\left[\n_1(x);\n_1(y)\right]=-i\delta^{\prime}(x-y)$.   
We then have the following conjugated field and the momenta:
\[
\pi(x)= -\frac{1}{\sqrt{2}}(\n_1(x)-\n_2(x));~~~~  
\d_x\phi(x)= \frac{1}{\sqrt{2}}(\n_1(x)+\n_2(x))
\]
In terms of these variables the Hamiltonian takes the following form: 
\be
H=\frac{1}{2}u(\l)\left[ (1/\xi)\pi^2(x)+ \xi(\d\phi(x))^2\right]
 = \frac{1}{2}u(\l)\left[ \hat{\pi}^2(x)+ (\d\hat{\phi}(x))^2\right], 
\label{h}
\ee
where 
\be
\pi(x)=\sqrt{\xi}~\hat{\pi}(x),~~~ 
\phi(x)=(1/\sqrt{\xi})\hat{\phi}(x).  
\label{canon}
\ee
The last equation (\ref{canon}) is nothing else as the canonical transformation,  
which is equivalent to the Bogoliubov transformation for the original operators 
$\rho_{1,2}(p)$. Next to establish the expressions for Fermions one should use 
the commutational relations $\left[a^{+}(x);\rho_{1}(p)\right]=-e^{ipx}a^{+}(x)$
and the same for $c^{+}(x)$. Note that these last relations were obtained using the 
expression with original fermions: 
$\rho_{1}(p)=\int dy e^{ipy}a^{+}(y)a(y)$. 
In this way we obtain the following expressions for fermionic operators:
\be
a^{+}(x)=K_1^{+}\frac{1}{\sqrt{2\pi\a}}
\exp\left(\frac{2\pi}{L}\sum_{p\neq0}\frac{\rho_1(p)}{p}e^{-ipx}e^{-\a|p|/2}\right)
=K_1^{+}\frac{1}{\sqrt{2\pi\a}}\exp\left(-i2\pi \tilde{N}_1(x)\right),
\label{fermions}
\ee
\[
c^{+}(x)=K_2^{+}\frac{1}{\sqrt{2\pi\a}}
\exp\left(-\frac{2\pi}{L}\sum_{p\neq0}\frac{\rho_2(p)}{p}e^{-ipx}e^{-\a|p|/2}\right)
=K_2^{+}\frac{1}{\sqrt{2\pi\a}}\exp\left(i2\pi \tilde{N}_2(x)\right),
\]
where the fields $\tilde{N}_{1,2}(x)$ are the analogs of the particle numbers 
with positions left to the point $x$  
and $K_{1,2}^{+}$ are 
the Klein factors - the operators 
which commute with the operators $\rho_{1,2}(p)$  
and create the single particle at the right (left) Fermi -points when 
acting on the ground state. The parameter $\a\to 0$ is introduced to 
perform the ultraviolet cutoff. The operators (\ref{fermions}) have the correct 
anticommutational relations.

It is assumed that there is an equality between the matrix elements 
of the specific 1D model for the low-energy states and the matrix elements 
of the corresponding operators over the corresponding eigenstates 
in the Luttinger liquid theory. 
The mapping between the two models is characterized by a single parameter 
$\xi$ which should be the same for both models in a sense of the 
equation (\ref{finite}) 
i.e. the constant 
$\xi$ in the effective Luttinger model is taken from the expression (\ref{finite}) 
for the original model. 
Thus the critical exponents in the asymptotics 
of the correlators for the 1D models are expressed in terms of the single 
parameter $\xi$ i.e. exhibit the universal behaviour. 
The constant $\xi$ determines the asymptotic behaviour 
of the correlators in the original model according to the prescriptions for the 
Luttinger model. 
This hypothesis is confirmed by the fact that the expression 
(\ref{finite}) is valid for the original model with the corresponding true 
speed of sound $v=u(\l)$. 
This was proved both for the continuous Bose or Fermi liquids and for 
the XXZ- spin chain \cite{H}.

\vspace{0.2in}

{\bf 3. Lowest formfactors.} 

\vspace{0.2in}

Let us start with the derivation of the scaling relations for the lowest 
formfactors for the XXZ- quantum spin chain: 
\[
H=\sum_{i=1}^{L}\left(\s^{x}_i\s^{x}_{i+1}+\s^{y}_i\s^{y}_{i+1}
+\D\s^{z}_i\s^{z}_{i+1}\right),  
\]
where the sites $L+1$ and $1$ are coincide. First, let us establish the relation 
for the formfactors of $\s_i^{\pm}$- operators. The relations for the other operators 
($\s_i^{z}$) as well as for the formfactors for the other 1D systems can be 
obtained in a similar way. 
Calculation of finite - size corrections to the energy of the ground state for 
the XXZ- spin chain (see for example \cite{Kar}) leads to the expression  
(\ref{finite}) and allows one to obtain the parameter $\xi$ which leads to 
the predictions of critical indices according to the conformal field theory. 
The calculation gives the value $\xi=2(\pi-\eta)/\pi$, where the parameter $\eta$ 
is connected with the anisotropy parameter of the XXZ - chain as 
$\Delta=\cos(\eta)$. 
Using the Jordan-Wigner transformation $\sigma^{+}_x=a^{+}_x\exp(i\pi N(x))$,
where $a^{+}_x$ stands for the ``original'' lattice fermionic operator, 
$N(x)=\sum_{l=1}^{x-1}n_l$ ($n_l=a_l^{+}a_l$),   
and performing the obvious substitutions $N(x)\rightarrow x/2+N_1(x)+N_2(x)$ and
$a^{+}_x\rightarrow e^{-ip_Fx}a^{+}(x)+e^{ip_Fx}c^{+}(x),~~p_F=\pi/2$,  
we obtain after the canonical transformation (\ref{canon})
the expression for the leading term in the asympotics of the correlator for the XXZ -chain:
\be
\la\sigma^{+}_x\sigma^{-}_0\ra=\la a_x^{+}e^{i\pi N(x)}a_0\ra
\sim(-1)^x
\la0|e^{-i\pi\sqrt{\xi}(\tilde{N}_1(x)-\tilde{N}_2(x))}
e^{i\pi\sqrt{\xi}(\tilde{N}_1(0)-\tilde{N}_2(0))} 
|0\ra,
\label{leading}
\ee
where $\tilde{N}_{1,2}(x)$ - are corresponds to the free fields 
$\hat{\pi}(x)$, $\hat{\phi}(x)$, obtained after the transformation (\ref{canon}). 
To these operators correspond the new operators $\rho_{1,2}(p)$ and the new 
fermionic operators (quasiparticles). 
Averaging the product of exponents in bosonic operators 
for the expression (\ref{leading}) and using the properties of $\rho_{1,2}(p)$, 
$\la\rho_1(-p)\rho_1(p)\ra=\frac{pL}{2\pi}\theta(p)$ and 
$\la\rho_2(p)\rho_2(-p)\ra=\frac{pL}{2\pi}\theta(p)$, 
we get for the correlation function 
$G(x)=\la0|\sigma^{+}_{i+x}\sigma^{-}_{i}|0\ra$ the following sum in 
the exponent: 
\[
C~\exp\left(~\frac{\xi}{4}~\sum_{n=1}^{\infty}\frac{1}{n}e^{in(2\pi x/L)}
+ h.c.\right), 
\]
where $C$ - is some constant. Then using the formula   
$\sum_{n=1}^{\infty}\frac{1}{n}z^n=-\ln(1-z)$ and substituting the value
$\xi=2(\pi-\eta)/\pi$ we obtain the following 
expression for the XXZ - chain: 
\be
G(x)=C_0\frac{(-1)^x}{\left(L\sin(\frac{\pi x}{L})\right)^{\alpha}},
~~~\alpha=\frac{\xi}{2}=\frac{\pi-\eta}{\pi} ~~~(x>>1).
\label{coas}
\ee
Thus, although bosonization, which deals with the low-energy effective theory, 
is not able to predict the constant before the asymptotics, the critical exponent 
and the functional form are predicted in accordance with conformal field theory.

From the relation (\ref{leading}) it follows the following equation: 
\be
\s^{-}_{0}=C^{\prime}K_{1}e^{i\pi\sqrt{\xi}(\Nt_1(0)-\Nt_2(0))}+\ldots.  
\label{operator}
\ee
where $C^{\prime}$ is some constant and the dots stand for the different 
(subleading) operators. Eq.(\ref{operator}) should be understood in a 
sense of the correspondence between the XXZ and the Luttinger liquid 
models: the formfactors for the corresponding states should be equal 
to each other. In particular for the ground states of $M$ ($|t\ra$) and 
$M-1$ ($|\lambda\ra$) particles (up-spins) for the XXZ spin chain 
we obtain: 
\be
\la\lambda|\s^{-}_{0}|t\ra=
C^{\prime}\la0|e^{i\pi\sqrt{\xi}(\Nt_1(0)-\Nt_2(0))}|0\ra.  
\label{lowest}
\ee
Calculating the average at the right-hand side of this equation 
we obtain $C^{\prime}=C(L/2\pi\a)^{\xi/4}$, where 
$C=\la\lambda|\s^{-}_{0}|t\ra$ is the value of the lowest formfactor. 
Next calculating the correlator $G(x)$ in the framework of the 
Luttinger model using the equation (\ref{operator}) we obtain the 
following result: 
\be
G(x)=(C^{\prime})^{2}\la0|e^{-i\pi\sqrt{\xi}(\tilde{N}_1(x)-\tilde{N}_2(x))}
e^{i\pi\sqrt{\xi}(\tilde{N}_1(0)-\tilde{N}_2(0))}|0\ra= 
\frac{C^2}{(2\sin(\pi x/L)^{\xi/2}}. 
\label{corr}
\ee
Note that the dependence of $G(x)$ on the parameter $\a$ is cancelled 
due to the dependence of the constant $C^{\prime}=C(L/2\pi\a)^{\xi/4}$ 
on $\a$ which indicates the independence of the scaling relations on 
the details of physics at high momenta which is model-dependent.  
Comparing the equation (\ref{corr}) with the equation (\ref{coas}) we 
obtain the desired relation between the lowest formfactor $C$ and 
the prefactor $C_0$: 
\be
C^2=\left(\frac{2}{L}\right)^{\xi/2}C_0.
\label{scaling}
\ee
For the case of the XX- spin chain ($\xi=1$) the relation between the 
formfactor and the prefactor (\ref{scaling}) coincides with the relation 
obtained in Ref.\cite{O} using the completely different method. 
Thus we see that while the prefactors are not the universal quantities 
the relations between them and the corresponding formfactors are universal.

Let us consider the density-density correlator for the XXZ- spin chain, 
namely $\Pi(x)=\la\s^{z}_{x}\s^{z}_{0}\ra$. 
Substituting the expressions for the original lattice Jordan-Wigner 
fermions into the density operator $n_x=a^{+}_{x}a_x$, we obtain the 
general expression for the density in terms of the Luttinger liquid 
operators: 
\be
n_x=\frac{1}{\sqrt{\pi}}\frac{1}{\sqrt{\xi}}\d_{x}\hat{\phi}(x)+ 
C^{\prime}_{1}e^{-i2p_{F}x}K_1^{+}K_{2}e^{-i2\pi(1/\sqrt{\xi})
\left(\Nt_1(x)+\Nt_2(x)\right)}+h.c.+\ldots,  
\label{oper}
\ee
where $p_F=\pi/2$, the operators $\Nt_{1,2}(x)$ correspond to the new 
fields $\hat{\pi}(x),\hat{\phi}(x)$ and the dots stand for the operators,
corresponding to the higher order terms in the expansion of the correlator. 
Consider the matrix element $\la t^{\prime}|n_0|t\ra=C_1$ , where $|t\ra$ -
is the ground state of the XXZ- spin chain and $|t^{\prime}\ra$ is the 
eigenstate which is obtained from the ground state by adding one particle 
at the right Fermi- point and removing one particle from the left Fermi- 
point. Taking the corresponding matrix element for both sides of the equation 
(\ref{oper}) we obtain: 
\be
\la t^{\prime}|n_0|t\ra=C_1=C_1^{\prime}
\la0|e^{-i2\pi(1/\sqrt{\xi})\left(\Nt_1(0)+\Nt_2(0)\right)}|0\ra.  
\label{me}
\ee
Calculating the average at the right- hand side side of Eq.(\ref{me}) 
we obtain the constant $C_1^{\prime}=C_1(L/2\pi\a)^{1/\xi}$. 
Calculating the correlator $\Pi(x)$ in the framework of the Luttinger 
liquid theory using the equation (\ref{oper}), we obtain the expression 
for the first two terms in the expansion in the form: 
\be 
\Pi(x)=\frac{1}{2\xi(L\sin(\pi x/L))^2}+ e^{i2p_{F}x}
\frac{C_1^2}{\left(2\sin(\pi x/L)\right)^{2/\xi}}+h.c.. 
\label{dens}
\ee
Again the parameter $\a$ drops out of this equation. 
Comparing this equation with the general expression for the correlator: 
\be 
\Pi(x)=\frac{1}{2\xi(L\sin(\pi x/L))^2}+ 
\frac{C_{10}\cos(2p_{F}x)}{\left(L\sin(\pi x/L)\right)^{2/\xi}}+\ldots,  
\label{density}
\ee
we obtain the scaling relation: 
\be
C_1^2=\frac{1}{2}\left(\frac{2}{L}\right)^{2/\xi}C_{10}. 
\label{scaldens}
\ee
The equation (\ref{scaldens}) was proved explicitly by direct 
computations for the XXZ spin chain in Ref.\cite{S1}. 

The examples presented above allow one to conclude the following. 
First, it is clear that in the same way the similar relations 
can be obtained for the other models of the 1D quantum liquids,  
in particular for the correlators of the continuous Bose- or 
Fermi- liquids. Second, the scaling relations can be easily 
generalized to the case of the lowest formfactors corresponding 
to the higher terms in the asymptotics of the correlators. 
The corresponding states are the states obtained from the ground 
state by moving an arbitrary number ($m$) of particles from the 
left to the right Fermi- points, which corresponds to the operators 
containing an additional powers of the operator $(a^{+}(x)c(x))$: 
$(a^{+}c)^m$. 
It is clear that in all the cases the results have the same form 
as the equations (\ref{scaling}), (\ref{scaldens}) with the 
corresponding critical exponent $\a(m)$. 
An explicit expressions for the correlators and the corresponding 
scaling relations for the 1D continuous Bose- and Fermi- liquids 
are presented in Ref.\cite{G}. 
For completeness we present these results in the Appendix.  
Here we would like to stress once more 
that these results of Ref.\cite{G} can be equally well obtained 
in the framework of the bosonization approach.

\vspace{0.2in}

{\bf 4. Particle-hole formfactors.} 

\vspace{0.2in}

Let us consider the calculation of the low-energy particle-hole 
formfactors in the framework of the bosonization approach. 
As an example consider the formfactor of the operator $\s_0^{-}$ 
for the XXZ spin chain. Suppose we have the eigenstate 
$\la\lambda(p_i,q_i)|$ obtained from the ground state by creating 
the holes with the momenta $q_i$ and the particles with the 
momenta $p_i$ ($i=1,\ldots n$) located in the vicinity of the 
right Fermi- point.
The formfactor corresponding to 
this state has the following representation in terms of Luttinger 
liquid matrix element: 
\be
\la\lambda(p_i,q_i)|\s_0^{-}|t\ra=
C^{\prime}\la{p_i,q_i}|e^{i\pi\sqrt{\xi}(\Nt_1(0)-\Nt_2(0))}|0\ra, 
\label{ph}
\ee
where at the right-hand side 
$p_i>0$ and $q_i\leq 0$ are the positions of the particles 
and the holes at the first branch of the Luttinger liquid 
(i.e. correspond to the operators $a^{+}(x)$, $a(x)$) and the 
constant $C^{\prime}$ was introduced in Eq.(\ref{operator}). 
Calculating the average at the right-hand side of the 
equation (\ref{ph}) we obtain the following result: 
\be 
\la\lambda(p_i,q_i)|\s_0^{-}|t\ra= 
C\la{p_i,q_i}|e^{a\frac{2\pi}{L}\sum_{p>0}\frac{\rho_1(p)}{p}}|0\ra, 
~~~~a=-\sqrt{\xi}/2,
\label{Lph}
\ee
where $C$ is the value of the lowest formfactor 
$C=\la\l|\s_0^{-}|t\ra$. 
The average at the right- hand side of the equation (\ref{Lph}) 
was calculated in Ref.\cite{A}: 
\be 
\la{p_i,q_i}|e^{a\frac{2\pi}{L}\sum_{p>0}\frac{\rho_1(p)}{p}}|0\ra
=F(p_i,q_i)=\det_{ij}\left(\frac{1}{p_i-q_j}\right)
\prod_{i=1}^{n}f^{+}(p_i)\prod_{i=1}^{n}f^{-}(q_i),
\label{F1}
\ee
where
\[
f^{+}(p)=\frac{\Gamma(p+a)}{\Gamma(p)\Gamma(a)}, ~~~~
f^{-}(q)=\frac{\Gamma(1-q-a)}{\Gamma(1-q)\Gamma(1-a)}. 
\]
In the equation (\ref{F1}) $p_i$ and $q_i$ are assumed to be 
integers (corresponding to the momenta $2\pi p_i/L$ and 
$2\pi q_i/L$) and $p_i>0$, $q_i\leq 0$.

Thus we have calculated the particle-hole formfactor in the form 
$\la\lambda(p_i,q_i)|\s_0^{-}|t\ra=CF(p_i,q_i)$, and have shown 
that the constant $C$ is in fact the lowest formfactor, which is 
not clear within the approach of Ref.\cite{G}. 
The same analysis can be performed for the formfactor 
$\la t^{\prime}(p_i,q_i)|n_0|t\ra=C_{1}F(p_i,q_i)$, where the 
constant $a$ in Eq.(\ref{F1}) is now equals $a=1/\sqrt{\xi}$, 
and the formfactors for the continuous Bose- and Fermi- liquids. 
Note that the dependence on the particles and holes momenta 
(\ref{F1}) can be easily obtained from the expressions for the 
formfactors in the case of the XX- spin chain in the form of 
the Cauchy determinant in Ref.\cite{O} ($\xi=1$, $a=-1/2$). 
We considered the formfactors corresponding to the particles 
and the holes located near the right Fermi point. 
The same analysis can be performed for the particles 
and the holes located near the left Fermi point. 
The total formfactor is a product of these two terms.

To sum up the formfactor series for the correlators in the 
framework of the approach of Ref.\cite{S2} (where the dependence 
on the particle and hole positions was found for the formfactors 
of the XXZ spin chain) it is useful to have the formula for the 
sum: 
\be
\sum_{n}\sum_{p_i>0,q_i\leq0}|F(p_i,q_i)|^2 e^{i(p-q)2\pi x/L}=
\frac{1}{(1-e^{i2\pi x/L})^{a^2}}, 
\label{sum}
\ee
where $p=\sum_{i=1}^{n}p_i$, $q=\sum_{i=1}^{n}q_i$ and 
$p_i$, $q_i$ are integers. 
The result (\ref{sum}) can be easily obtained with the help of 
the following correlator: 
\[
G_{a}(x)=\la0|
e^{-a\frac{2\pi}{L}\sum_{p<0}\frac{\rho_1(p)}{p}e^{-ipx}} 
e^{a\frac{2\pi}{L}\sum_{p>0}\frac{\rho_1(p)}{p}}|0\ra= 
\frac{1}{(1-e^{i2\pi x/L})^{a^2}}, 
\]
which can be calculated either with the help of inserting 
of the complete set of intermediate states or by means of the 
standard formulas used in the bosonization approach. 
Thus the two sides of the equation (\ref{sum}) should be equal 
to each other. 

Finally, it would be interesting to verify the sum rule for 
the function $F(p_i,q_i)$ which can be obtained from the 
equation (\ref{sum}) by means of expanding it into the 
Fourier series: 
\[
\sum_{n}\sum_{p_i,q_i,p-q=m}|F(p_i,q_i)|^2= 
\frac{\Gamma(a^2+m)}{\Gamma(m+1)\Gamma(a^2)}, 
\]
where $p=\sum_{i=1}^{n}p_i$, $q=\sum_{i=1}^{n}q_i$.

In conclusion, 
for various one-dimensional quantum liquids in the framework of the Luttinger model 
(bosonization) we established the relations between the prefactors of the 
correlators and the formfactors of the corresponding local operators. 
The derivation of these relations in the framework 
of the bosonization procedure allows one to substantiate the prediction for the 
formfactors corresponding to the low-lying particle-hole excitations. 
Let us stress once more that the relations of the type (\ref{Lph}) can be 
obtained only in the framework of the bosonization procedure, since only within 
this approach one can see that the constant $C$ in Eq.(\ref{Lph}) is in fact 
the lowest formfactor. We also obtained   
the formulas for the summation over the particle-hole states corresponding to the 
power-law asymptotics of the correlators.

\vspace{0.2in}

{\bf Appendix.} 

\vspace{0.2in} 

Here we present without derivation the scaling relations for the 
formfactors for the continuous Bose- and Fermi- liquids and the XXZ- 
spin chain including the relations for the higher order terms 
in the asymptotics of the correlators. The results for the 
Bose-liquid have exactly the same form as the results for the 
XXZ- spin chain. We denote by $\phi(x)$ ($\psi(x)$) the fields 
corresponding to the Bose (Fermi) particles and by $\rho(x)$ the 
density operator. The general expressions 
for the expansions of the Bose- field and the density operators are: 
\[
\phi(x)=\sum_{m}C_{m}^{\prime}e^{-i2p_{F}mx}
e^{i\pi\sqrt{\xi}(\Nt_1(x)-\Nt_2(x))} 
e^{-i2\pi m(1/\sqrt{\xi})(\Nt_1(x)+\Nt_2(x))}, 
\]
\[
\psi(x)=\sum_{m}B_{m}^{\prime}e^{i(2m+1)p_{F}x}
e^{i\pi\sqrt{\xi}(\Nt_1(x)-\Nt_2(x))} 
e^{i\pi(2m+1)(1/\sqrt{\xi})(\Nt_1(x)+\Nt_2(x))}, 
\]
\[
\rho(x)=\frac{1}{\sqrt{\pi}}\frac{1}{\sqrt{\xi}}\d_{x}\hat{\phi}(x)+ 
\sum_{m\neq0}A_{m}^{\prime}e^{-i2p_{F}mx} 
e^{-i2\pi m(1/\sqrt{\xi})(\Nt_1(x)+\Nt_2(x))},
\]
where $p_F=\pi\rho_0$ is the Fermi- momentum (here we omit the 
Klein factors, for the correlators one should take the averages 
of the terms, corresponding to the same harmonics (same $m$)). 
Exactly the same expressions hold for the operator $\s^{-}_x$ and 
$\s^{z}_x$ for the XXZ- spin chain ($p_F=\pi/2$) 
The general expressions for the equal-time correlators have the form: 
\[
G_B(x)=\la\phi^{+}(x)\phi(0)\ra= \sum_{m\geq0}C_{m}
\frac{\cos(2p_{F}mx)}{\left(L\sin(\pi x/L)\right)^{\xi/2+m^{2}(2/\xi)}}, 
\]
\[
G_F(x)=\la\psi^{+}(x)\psi(0)\ra= 
\sum_{m\geq0}B_{m}
\frac{\sin((2m+1)p_{F}x)}{\left(L\sin(\pi x/L)\right)^{\xi/2+(2m+1)^2/2\xi}}, 
\]
\[
\Pi(x)=\la\rho(x)\rho(0)\ra=\rho_0^2+\frac{1}{2\xi(L\sin(\pi x/L))^2}+
\sum_{m\geq 1}A_{m}
\frac{\cos(2p_{F}mx)}{\left(L\sin(\pi x/L)\right)^{(2/\xi)m^2}}. 
\]
The expressions for $G_B(x)$ and $\Pi(x)$ also hold for the XXZ- spin 
chain (the correlators $G(x)$ and $\Pi(x)$). Then the scaling 
relations for the lowest formfactors have the following form: 
\[
|\la\l(m)|\phi(0)|t\ra|^2=\frac{(-1)^{m}C_{m}}{2-\delta_{0,m}}
\left(\frac{2}{L}\right)^{\xi/2+m^{2}(2/\xi)}, 
\]
\[
|\la\l(m)|\psi(0)|t\ra|^2=\frac{(-1)^m B_{m}}{2}
\left(\frac{2}{L}\right)^{\xi/2+(2m+1)^2/2\xi}, 
\]
\[
|\la t(m)|\rho(0)|t\ra|^2=\frac{A_{m}}{2}
\left(\frac{2}{L}\right)^{(2/\xi)m^2}, 
\]
where $|\l(m)\ra$ is the eigenstate with the number of particles 
equal to $M-1$ and with $m$ particles removed from the right and 
created at the left Fermi- point, and $|t\ra$ is the ground state 
of $M$ particles ($\rho_0=M/L$). For the case of the XXZ- spin 
chain the particles correspond to the up-spins and we have the 
same formulas as the formulas for the bosons 
($\phi(0)\rightarrow\s_0^{-}$, $\rho(0)\rightarrow\s_0^{z}$). 
The factors $(-1)^m$ in the last equations appear in the process 
of the calculations of the corresponding averages for the 
correlators at $m\neq0$.


\vspace{0.3in}

\begin{thebibliography}{99}

\bibitem{LP}
A.Luther, I.Peschel, Phys.Rev.B 9 (1974) 2911;~ Phys.Rev.B 12 (1975) 3908.

\bibitem{H}
F.D.M.Haldane, Phys.Rev.Lett. 47 (1981) 1840;~ J.Phys.C 14 (1981) 2585~;
Phys.Rev.Lett. 45 (1980) 1358.  

\bibitem{C}
J.Cardy, Nucl.Phys. B 270 (1986) 186. 

\bibitem{MZ}
A.Mironov, A.Zabrodin, Int.J.Mod.Phys. A7 (1992) 3885. 

\bibitem{S1}
N.Kitanine, K.K.Kozlowski, J.M.Maillet, N.A.Slavnov, V.Terras, \\ 
J.Stat.Mech.:Theory and Experiment 04 (2009) P04003.  

\bibitem{G}
A.Shashi, L.I.Glazman, J.S.Caux, A.Imambekov, arXiv:1103.4176 [cond-mat], 
arXiv:1010.2268 [cond-mat]. 

\bibitem{S2}
N.Kitanine, K.K.Kozlowski, J.M.Maillet, N.A.Slavnov, V.Terras, 
arXive:1003.4557 [math-ph].  

\bibitem{ML}
D.Mattis, E.Lieb, J.Math.Phys. 6 (1965) 304. 

\bibitem{Kar}
H.J.de Vega, M.Karowski, Nucl.Phys.B 285 (1987) 619; F.Woynarovich, H.P.Eckle, 
J.Phys.A 20 (1987) L97; M.Karowski, Nucl.Phys.B 300 (1988) 473; 
F.C.Alcaraz, M.N.Barber, T.M.Batchelor, Phys.Rev.Lett. 58 (1987) 771. 

\bibitem{O}
A.A.Ovchinnikov, J.Phys.:Condens.Matter 16 (2004) 3147.  

\bibitem{A} 
E.Bettelheim, A.G.Abanov, P.Wiegmann, J.Phys.A 40 (2007) F193. 



\end{thebibliography}
\end{document}